\documentstyle[preprint,aps]{revtex}

\begin{document}
 \draft

 \title{Chemical Control of Spin Chirality in (Nd$_{1-x}$Dy$_{x}$)$_{2}$Mo$_{2}$O$_{7}$ }
 \author{S. Iguchi$^{1}$, D. Hashimoto$^{1}$, Y. Taguchi$^{2}$, and Y. Tokura$^{1,3,4}$}

 \address{$^{1}$ Department of Applied Physics, University of Tokyo, 
Tokyo 113-8656, Japan}
 \address{$^{2}$ Institute for Materials Research, Tohoku University, Sendai 980-8577, Japan}
 \address{$^{3}$ Spin Superstructure Project (SSS), ERATO, 
Japan Science and Technology Corporation (JST), Tsukuba 305-8562, Japan}
 \address{$^{4}$ Correlated Electron Research Center (CERC), 
National Institute of Advanced Industrial Science and Technology 
(AIST), Tsukuba 305-8562, Japan}

 \date{received\hspace*{3cm}}
 \maketitle

 \begin{abstract}

Magnetization and Hall resistivity 
 have been investigated for single crystals of
Dy-doped Nd$_{2}$Mo$_{2}$O$_{7}$.
A sharp decrease of the Hall resistivity upon a metamagnetic transition, 
perhaps associated with magnetic-field ($H$) induced flop of Dy$^{3+}$ moments, 
has been observed in the Dy-doped crystals only for 
$H \parallel [111]$ direction.
In addition, the sign of the Hall resistivity at  a high  field, 
both for $H \parallel [100]$
and for $H \parallel [111]$,
changes with the Dy-doping.
These results are explained in terms of  the sign change of Mo spin chirality 
that is controlled by the Dy$^{3+}$ moments with a different sign of $f$-$d$
interaction from the Nd$^{3+}$ moments.

 \end{abstract}

 \bigskip
 \pacs{PACS number(s):75.47.-m, 75.30.Mb, 75.50.Cc}
 \narrowtext

Transport properties in strongly correlated systems are largely affected by 
the configuration of  spin background.
Typical example is colossal magnetoresistance phenomena in manganites~\cite{Tokura99},
where magnitude of transfer interaction, and hence longitudinal conductivity, 
is controlled by the degree of ferromagnetic spin correlation~\cite{Anderson55}. 
Recently, it has  been argued~\cite{Ye99,Ohgushi00,MOnoda02,SOnoda03} that 
 non-trivial spin texture can give rise to quantal phase  of the electron
 transfer interaction,
and that the phase, which acts as a fictitious magnetic
field, manifests itself as  transverse conductivity.
Experimental evidences 
for this mechanism of anomalous Hall effect (Berry phase mechanism)
 have been accumulated for various kinds of 
oxide materials~\cite{Matl98,Chun00,Taguchi01,Yanagihara02,Taguchi03}.
Among them, unconventional behaviors, such as temperature-variation and
anisotropic magnetic field-dependence,
 of the anomalous Hall effect in
Nd$_2$Mo$_2$O$_7$ with pyrochlore structure~\cite{Taguchi01,Taguchi03,Kageyama01}
 are considered as strong evidences for the Berry phase mechanism
of anomalous Hall effect,
although they are not interpreted as such in some literature~\cite{Kageyama01}.

The unique behaviors of the anomalous Hall effect in Nd$_2$Mo$_2$O$_7$
arise from the complex magnetic structure~\cite{Taguchi01,Yasui01} 
realized on that particular pyrochlore lattice.
The lattice consists of two sublattices of Nd-site and Mo-site, both of which
form corner-sharing networks of tetrahedra. 
One sublattice is obtained by 
displacing the other one by half a lattice constant along a crystallographic
axis. The localized Nd-moment has Ising-like anisotropy along the local 
$\langle 111 \rangle$ axis which differs
from site to site, and the anisotropy of the Nd moment system is transmitted via
$f$-$d$ interaction to the Mo spin system, with which conduction electrons 
interact via Hund's-rule coupling.
Sign reversal of the Hall resistivity at high fields applied along the 
[111] direction  is explained~\cite{Taguchi03} in terms of 
field-induced flipping of Nd moments and consequent
sign change of  fictitious magnetic field that penetrates a Mo-tetrahedron.
On the other hand, a rather conventional behavior of 
anomalous Hall effect in Gd$_2$Mo$_2$O$_7$ 
is ascribed~\cite{Taguchi03}  to the isotropic 
nature of Gd$^{3+}$ moment without orbital angular momentum, and to the resultant
absence of spin chirality in the Mo spin system.
Therefore, the anisotropy of the
magnetic moment of the rare earth ion  is a key ingredient in 
this mechanism of anomalous Hall effect.
In this paper, we fully exploit the fact that the spin chirality of the Mo system
is controlled by the moment of rare earth species, and report 
on the variation of the Hall resistivity with the partial 
Dy-substitution on the Nd-site.
The Dy$^{3+}$ with strong Ising-anisotropy~\cite{Ramirez99,Fukazawa02,Sakakibara03} 
has a larger moment ($\approx 10 \mu_{B}$) 
than the Nd$^{3+}$ ($\approx 2.3 \mu_{B}$), 
and hence is more amenable to an applied
field. Even more important is the fact that 
the Dy-Mo interaction  is ferromagnetic  as opposed 
to antiferromagnetic Nd-Mo interaction, as we will see.
Therefore, Mo spins near the Dy moments are tilted in a different way
from the undoped sample, which enables us to chemically  control the
spin-chirality related phenomena in terms of Dy-doping into 
Nd$_{2}$Mo$_{2}$O$_{7}$.

All the samples used in this study were single crystals prepared by a 
floating-zone method in Ar atmosphere.  
Powder X-ray diffraction measurements  indicated that 
the samples were of single phase.
Hall resistivity  and magnetization measurements were done on 
the same sample for each Dy-concentration and
sample orientation.
The Hall resistivity was measured by
using the conventional four-probe method.
We extracted the odd component of Hall voltage with respect to the reversal of field
direction in order to eliminate the
longitudinal voltage drop due to the asymmetry of Hall voltage probes.

In Figs.1(a-c), we show the magnetization process (field-decreasing run) of 
(Nd$_{1-x}$Dy$_x$)$_2$Mo$_2$O$_7$
with $x$=0 (at 1.6 K), $x$=0.2 and $x$=0.35 (at 2 K), whose Curie
temperatures determined from
ac-susceptibility measurements are 89 K, 80 K, and 76 K, respectively. 
The results of the undoped ($x$=0) sample were reproduced from 
ref.~\cite{Taguchi03}.
 Applied field directions are along [100] (broken lines) and 
[111] (closed lines).
The magnetization due to the Mo spin system in this temperature range is
about 1.4$\mu_{\rm B}$/(Nd,Dy)MoO$_{3.5}$~\cite{Taguchi01}, and is 
indicated by a horizontal dotted line.
The magnetization process of the $x$=0 crystal
can be viewed~\cite{Taguchi03} as gradual flipping process of Nd moments
that are coupled antiferromagnetically with Mo spins at zero field.  
The partial substitution of Dy moments produces an almost parallel shift of 
the magnetization curves toward higher values.
The Dy moments in the pyrochlore lattice
are well known~\cite{Ramirez99,Fukazawa02,Sakakibara03} to behave like
$\langle 111 \rangle$ Ising moment
with larger magnitude ($\approx$ 10$\mu_{\rm B}$)
than that of Nd moment ($\approx$ 2.3$\mu_{\rm B}$).
The Dy-doping dependence of the magnetization values at low 
(0 T from the field-decreasing run) and high (14 T) field are shown in Fig.2(a).
The increased magnetization  at zero-field upon the partial Dy 
substitution indicates that the interaction between
Dy moments and Mo spin system is basically ferromagnetic. 
As indicated by upward arrows in Figs.1(b) and (c), a metamagnetic
anomaly is clearly discerned at about 3 T in the Dy-doped samples
when the field is applied along the [111] direction.
We ascribe this anomaly to the collective flipping  of minority part of  Dy moments 
that point antiparallel to the net magnetization direction, as we will later 
discuss in detail.
There is no trace of such a metamagnetic  anomaly either 
in the case of $H \parallel$ [100] direction for any $x$ or in the $x$=0 crystal
for any field direction.

In Figs.1(d-f), Hall resistivity($\rho_{H}$) is plotted against magnetic field
($H$) for the samples with $x$=0, 0.2, and 0.35.
The Hall resistivity of the undoped sample undergoes a sign change when 
$H \parallel$ [111], while gradually approaching zero for $H \parallel$ [100]. 
This behavior is in accord with the prediction by 
 the Berry phase theory,
and has been interpreted~\cite{Taguchi03} in terms of field-induced sign reversal  of the Mo spin chirality.
The $x$=0.2 Dy-doped crystal also exhibits a similar feature with 
the $x$=0 crystal. 
For $x$=0.35, by contrast, the Hall resistivity shows a qualitatively 
different behavior in the high field region: 
The sign of $\rho_{H}$ changes from positive to negative 
when $H \parallel$ [100], 
while for $H \parallel$ [111] the sign remains positive.
On the basis of chirality mechanism, this result implies that the Dy moments 
interacting ferromagnetically with the Mo spins may alter the sign of the
Berry phase of Mo conduction electrons in an opposite way to the case of the 
undoped Nd$_{2}$Mo$_{2}$O$_{7}$ in the high field region.
The sign change of the high-field(14 T) Hall resistivity as a function of Dy 
doping is summarized for the both field directions in the Fig.2(b).
In the low-doped region of $x \le$ 0.25, the Hall resistivity at 14 T is positive
or almost zero, while
it becomes negative for $x=$ 0.35.
Conversely, the Hall resistivity for $H \parallel$ [111] direction is negative
for $x \le$ 0.25, but becomes positive at $x=$  0.35.

What is also unique for the Dy-doped crystal is
a sudden decrease  of Hall resistivity associated 
with  the metamagnetic anomaly at about 3 T, as indicated
by downward arrows in Figs.1(e) and (f).
The feature shows up most clearly in the $x$=0.20 case, but is also discerned
 for $x$=0.35. 
To show the temperature dependence of the anomaly, we show in Figs.3(a) and (b)
the magnetization and Hall resistivity of the $x$=0.2 sample,
respectively, at various  temperatures
as a function of $H \parallel$ [111].
As the temperature is elevated from 2 K, 
the magnetic field at which the anomaly occurs
shifts gradually to lower field, and eventually disappears at about 50 K.
In the inset to Fig.3 is shown the relation 
between temperature and magnetic field
where the metamagnetic transition is observed.
The transition point is defined as the magnetic field where
$\frac{dM}{dH}$ value takes maximum at each temperature.

In the case of the undoped sample, the Nd moments form
umbrella structure, or \lq\lq 2-in 2-out\rq\rq \ 
structure~\cite{Taguchi01,Yasui01}, at 
a small enough field. The magnetic structure is depicted for
$H \parallel$ [111] configuration  in Fig.4(a),
where $H$ direction is upward.
Figures 4(b)-(e) are the cases where only one Dy is doped
into the single tetrahedron. (We neglect the cases where more than
one Dy ions are included in the tetrahedron, for simplicity.)
The configuration of the Dy moment is determined by the combined
effect of $f$-$d$ and $f$-$f$ interactions.
The former is ferromagnetic as we have already noted, and the
latter must be antiferromagnetic on the basis of the following 
argument.
If the $f$-$f$ interaction could be neglected,
 all the Dy moments would point to the same direction as 
the Mo spins or the magnetic field even at a small enough field.
In this case, the expected value of  the magnetization increase upon the
20 \% Dy doping at a small enough field would be
1.2$\mu_{\rm B}$/(Nd,Dy)MoO$_{3.5}$
, which is much larger than the observed value
of 0.6$\mu_{\rm B}$/(Nd,Dy)MoO$_{3.5}$ (see Fig.2(a)).
Therefore, the $f$-$f$ interaction cannot be neglected and has to
be taken into account.
If the Dy moment substituted for the site 4 points 
antiparallel to the net magnetization as depicted in Fig.4(e), 
the magnetization increase is expected to be
0.8 $\mu_{\rm B}$/(Nd,Dy)MoO$_{3.5}$ at 20 \% Dy-doping. 
Provided that the doped Dy moments always point antiparallel 
to the replaced original Nd moments in the 2-in 2-out configuration,
the anticipated variation of the spontaneous ($H$=0) magnetization
 is shown with a dotted line in Fig.2(a), which is  very close to the observed.
The configuration (e) is possible only when the Dy-Nd interaction
is antiferromagnetic. 
Note that the antiferromagnetic Dy-Nd interaction
is compatible with configurations (b)-(d)
that are made by the ferromagnetic Dy-Mo interaction.
When $H$ is increased, the Dy moments at the site 4
collectively flip so as to gain the Zeeman energy, at the expense of the 
Dy-Nd interaction energy. 
The collective flipping  of Dy moments corresponds to the metamagnetic
anomaly observed at around 3.5 T  at 2 K. 
The observed jump of magnetization in the 20\% 
Dy-doped sample is about 0.3-0.4 
$\mu_{\rm B}$/(Nd,Dy)MoO$_{3.5}$, which is in  good agreement
with the expected value of 0.33 $\mu_{\rm B}$/(Nd,Dy)MoO$_{3.5}$
for the spin-flip process (e)$\rightarrow$ (f) in Fig.4.
The interaction energy, or equivalently the molecular
field from the Nd moments,  which prevents  Dy moments from flipping, 
grows in magnitude as the ordered moment of Nd  increases.
This is the reason why the temperature dependence of the
metamagnetic anomaly (shown in the inset to Fig.3) is qualitatively 
similar to  the evolution of the ordered Nd moments.

In the case of $H \parallel$ [100] as depicted in Fig.4(n), 
by contrast, all the Dy moments seem to point toward the field direction.
This is because  this configuration 
is  compatible  with the antiferromagnetic Dy-Nd
as well as the ferromagnetic Dy-Mo interactions.
In fact,  the observed Dy-doping dependence  of magnetization for a small 
$H \parallel$ [100] is in
reasonable agreement with the expected value (a dashed line in Fig.2(a)),
which was derived again on the assumption that the doped Dy moment
always points antiparallel to the original Nd moment.
Therefore, all the Dy moments are already aligned along the field 
direction
at a small enough field,  which explains why no metamagnetic anomaly 
is observed in the magnetization curve for $H \parallel$ [100].

Keeping these features of the (Nd,Dy) moment configuration in mind,
we discuss the Hall resistivity anomaly upon 
the Dy-related metamagnetic transition
as well as the Dy-doping induced sign reversal of 
Hall resistivity at a high field. 
As we have argued, the moments which show field-induced flipping are
the Dy moments at site 4 (Fig.4(e)).
The magnetic structures of the tetrahedron right after the Dy
flipping are depicted in Figs.4(f) and (g).
For the purpose of simplicity, we assume that all the (Nd,Dy)-tetrahedra
which surround one Mo tetrahedron 
have the same magnetic structure.
Then, the fictitious field changes the sign upon flipping of Dy
moment at site 4. 
The sign of the fictitious magnetic field
that penetrates the Mo tetrahedron is indicated by \lq\lq $+$\rq\rq \
or \lq\lq $-$\rq\rq \ in the each panel of  Fig.4.
This gives rise to the sudden decrease of 
Hall resistivity.
At a high enough field (e.g. 14 T),  all the Nd and Dy moments 
point to the field direction to gain the Zeeman energy, as shown
in Figs.4(h)-(l).  
If we take into account the different sign of $f$-$d$ interaction 
between Nd-Mo and Dy-Mo, 
the sign of the fictitious magnetic field becomes positive in some
of the tetrahedra that include Dy ions, as opposed to negative sign
in the tetrahedra without Dy.
Therefore, as the Dy concentration is increased, 
the number of the tetrahedra that include Dy increases, and eventually
the sign of the
Hall resistivity at the high field changes from negative to positive.
On the other hand,  when the  strong magnetic field is applied 
along [100] direction, a tetrahedron that includes  Dy ion makes a 
negative-sign contribution to the Hall resistivity (Fig.4(p)).
Thus, the sign of the Hall resistivity changes from positive
to negative, just in the opposite way to the case of $H \parallel$ [111].

In summary, we have investigated the anisotropic magnetization and
Hall resistivity for a series of (Nd$_{1-x}$Dy$_{x}$)$_2$Mo$_2$O$_7$
single crystals.
A metamagnetic anomaly is observed in the magnetization curve
 for $H \parallel$ [111].
The Hall resistivity shows a sudden decrease upon the metamagnetic transition.
We attribute this anomaly to collective flipping of minority part
of Dy moments, and the decrease of Hall resistivity to the sign change of 
fictitious field  that penetrates Mo tetrahedra near the flipped Dy moments.
As the Dy concentration is increased, 
the sign of the Hall resistivity at a high enough field has changed from
negative to positive for $H \parallel$ [111], and vice versa for 
$H \parallel$ [100].
This is caused by the change of Mo spin tilting that is controlled by
the Dy moments with different sign of $f$-$d$ interaction from that of Nd.

We would like to thank N. Nagaosa, S. Onoda and N. Hanasaki
for enlightening discussions.
This work was in part supported by a Grant-In-Aid
for Scientific Research  from the MEXT, Japan.

\begin{figure}
\caption{Magnetic field dependence of (a-c)
 magnetization and (d-f) Hall resistivity of 
(Nd$_{1-x}$Dy$_x$)$_2$Mo$_2$O$_7$ crystals with $x$=0, 0.2, and 0.35 for
the magnetic-field ($H$) directions,
$H \parallel [100]$ and $H \parallel [111]$.
Arrows in (b,c) and (e,f) indicate the position of 
 the metamagnetic anomaly.
A horizontal broken line in (a-c) represents the contribution to the total
 magnetization from the  Mo spins alone.}
\end{figure}

\begin{figure}
 \caption{
Dy doping ($x$) dependence of (a) magnetization at 2 K and at 0 T
 and 14 T, and 
(b) Hall resistivity at 2 K and 14 T for
 (Nd$_{1-x}$Dy$_{x}$)$_{2}$Mo$_{2}$O$_{7}$ with $H \parallel [100]$ 
and $H \parallel [111]$ geometries.
Dotted lines are the $x$-dependence at 0T calculated by assuming that
the moment of
the doped Dy points the opposite direction to the original Nd moment on
the same site, as expemplified in (b)-(e) and (n) of Fig.4.
Dashed lines are for the figh-field case calculated by assuming that all
the Dy and Nd moments are aligned, while keeping the Ising anisotropy,
toward the $H$ direction, as shown in the right-most column (h)-(l), (o) and
(p) in Fig.4.
}
\end{figure}

\begin{figure}
 \caption{
(a) Magnetization and (b)  Hall resistivity 
of (Nd$_{1-x}$Dy$_{x}$)$_{2}$Mo$_{2}$O$_{7}$ with $x$=0.2 
at various temperatures are plotted
against applied magnetic field ($H$) along [111].  
The inset shows the relation between the field and the temperature  ($T$)
where the metamagnetic transition occurs.
}
\end{figure}

\begin{figure}
 \caption{The magnetic structure of rare-earth tetrahedra.
In the panels (a)-(l) ((m)-(p)), magnetic field ($H$) is applied along
[111] ([100]) direction.
Panels (a)-(e), (m), and (n) correspond to the cases with a small enough magnetic field while
panels (h)-(l), (o), and (p) to the cases with a high magnetic field.
Panels (f) and (g) represent the configuration right after the flipping
of Dy moments at site 4.
Nd  and Dy moments are represented by smaller  and larger arrows, respectively.
The \lq\lq +/- \rq\rq \ symbol indicates the sign of fictitious field that
penetrates Mo tetrahedra.
}
\end{figure}

\end{document}